\documentclass[twocolumn,prb,showpacs,preprintnumbers,amsmath]{revtex4}

\usepackage{graphicx} %Include figure files
\usepackage{epsfig}   %Include postscript files
\usepackage{bm}       %bold math
\usepackage[usenames]{color}

\begin{document}

\preprint{{\it J.~Chem.~Phys.}, submitted 11/12/03.}
% \preprint{JCP - \ddmmyyyydate \today~\xxivtime}

\title{An Empirical Charge Transfer Potential with \\ Correct
Dissociation Limits}

\author{Steven~M.~Valone}
%\email{smv@lanl.gov}
\affiliation{Materials Science and Technology Division, Los Alamos
National Laboratory,
Los Alamos, New Mexico 87545 and \\
Department of Physics and Astronomy, University of New Mexico,
Albuquerque, New Mexico 87131}

\author{Susan~R.~Atlas}
%\email{susie@sapphire.phys.unm.edu}
\affiliation{Center for Advanced Studies and
Department of Physics and Astronomy, \\ University of New Mexico,
Albuquerque, New Mexico 87131 \vspace*{.05in}}

\date{\today}

\begin{abstract}
  The empirical valence bond (EVB) method [J.~Chem.~Phys.~{\bf 52}, 1262
  (1970)] has always embodied charge transfer processes.  The mechanism
  of that behavior is examined here and recast for use as a new
  empirical potential energy surface for large-scale simulations.  A
  two-state model is explored.  The main features of the model are: (1)
  Explicit decomposition of the total system electron density is
  invoked; (2) The charge is defined through the density decomposition
  into constituent contributions; (3) The charge transfer behavior is
  controlled through the resonance energy matrix elements which cannot
  be ignored; and (4) A reference-state approach, similar in spirit to
  the EVB method, is used to define the resonance state energy
  contributions in terms of ``knowable'' quantities.  With equal
  validity, the new potential energy can be expressed as a {\it
  nonthermal} ensemble average with a nonlinear but analytical charge
  dependence in the occupation number.  Dissociation to neutral species
  for a gas-phase process is preserved.  A variant of constrained search
  density functional theory is advocated as the preferred way to define
  an energy for a given charge.
\end{abstract}

% \keywords{Suggested keywords} %Use showkeys class option if keyword
% display desired
\maketitle

\section{INTRODUCTION}
Charge transfer is ubiquitous in physical processes
affecting biological, chemical, and materials systems.  The
representation of charge transfer is of intense current interest
throughout the physical sciences.  A powerful concept in both modeling
and understanding how charges redistribute themselves during a
physical process is chemical potential
equalization.\cite{SAN51,PPLB,ParrPear,RSB,QEq,ES+} To apply chemical
potential equalization successfully, it is essential to use a
charge-dependent energy model which behaves correctly for all
configurations encountered in the process of interest.

Consider a diatomic molecule AB.  Atom A is assumed to be more
electropositive than atom B.  We are interested in charge
disproportionation reactions typified by
\begin{equation}
     {\rm A}^0{\rm B}^0 \leftrightarrow {\rm A}^{+q}{\rm B}^{-q}
     \label{I.0a}
\end{equation}
and
\begin{equation}
     {\rm A}_2 + {\rm B}_2 \leftrightarrow 2 \, {\rm A}^{+q}{\rm
     B}^{-q} \ , \label{I.0b}
\end{equation}
where the charge $q$ is not necessarily an integer.  Charge
disproportionation and transfer reactions are important in electron
transfer in biophysical systems,\cite{Glycine,HamSchif-ET,Sheu} enzyme
catalysis reactions,\cite{Page,HamSchif-Rev} and such processes as
electron-hole production and recombination in organic
semiconductors.\cite{Organic} For dimers (B = A), Eq.~(\ref{I.0a})
corresponds to broken charge-symmetry states.  Furthermore,
Eqs.~(\ref{I.0a}) and (\ref{I.0b}) have been the prototype reactions
for a wide variety of theoretical studies of chemical
bonding.\cite{ParrPear,ParrBarto,Nal1,Pearson,
NalParr,Nal2,hir,CioStef,Voth}

The most prevalent model potential for charged species is the
quadratic expansion of the classical
electrostatic potential,\cite{RSB,ES+,Imac,BKS,abinitFF}
\begin{eqnarray}
     E \approx E_0(R) + E_1(R) \, q + 1/2 \, (E_2(R) - V(R)) \, q^2 \ .
     \label{I.1}
\end{eqnarray}
The $E_i(R)$ are expansion coefficients which are atom-type dependent
and $V(R)$ is an interionic potential representing the classical
electrostatic contribution.  In general all of these functions depend
on the separation $R$.  The charge $q$ is usually, but not always,
independent of $R$.  $V(R)$ approaches a $1/R$ dependence at
sufficiently large $R$.  $E_0(R)$ contains the charge-independent
short-range and dispersion interactions between the atoms.  The other
expansion coefficients are frequently interpreted in terms of physical
quantities such as chemical potential and
hardness.\cite{ICM61,Klopman,RSB,QEq,ES+,ParrYang,
vSNG1,vSNG2,vSNG6,Mortier,Geerlings} The quadratic form has the virtue
of simplicity and works adequately when the range of $R$ is small
enough to prevent $q$ from changing appreciably.

\begin{figure}
\includegraphics[scale=0.55]{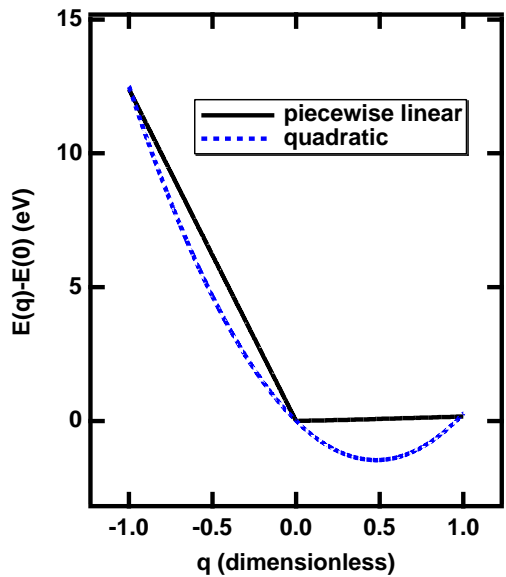}
\caption{\label{fig:V2U} Comparison of piecewise
linear and quadratic models of the charge dependence of the energy for
a diatomic.  The solid line corresponds to the energy of the {\it
isolated} atoms as a function of acquired charge $q$ (measured with
respect to charge transfer from the cation).  The dashed curve
corresponds to a quadratic representation of the charge dependence.
The energy minimizes at $q = 0$ for the piecewise linear model, but
not for the quadratic model.  At large separations, the piecewise
linear model is correct.  Traditionally the quadratic model is
considered to be more correct at finite separations.}
\end{figure}

If Eq.~(\ref{I.1}) is used at all separations, the atoms in the AB
molecule will remain ionic even at separations where they are supposed
to return to neutral states.  For those situations where charge
transfer does occur, alternative functional forms of charge dependence
need to be invoked.  For instance, Alavi et al.\cite{Imac} and
Grochowski and coworkers\cite{Try} use phenomenological switching
functions to effect charge transfer.  At very large separations,
though, it is known that the charge dependence becomes piecewise
linear (Fig.~\ref{fig:V2U}).\cite{PPLB,CioStef,Nal-KS}

Morales and Martinez\cite{ToddM} (hereafter referred to as MM)
conclude that the ``grand canonical'' (GC) approach\cite{PPLB} cannot
describe a realistic charge transfer process.  In the GC approach each
atom is considered to be an open system with respect to exchange of
energy and numbers of electrons.  They use a 3-state model with
integer charge resonances representing the states of the ensemble for
each atom.  For atom A for instance, the GC energy is expressed in the
form of an ensemble average
\begin{equation}
     E_{\rm A}^{\rm GC} \approx \omega_{\rm A^0} E_{\rm A^0} +
     \omega_{A^+} E_{\rm A^+} + \omega_{\rm A^-} E_{\rm A^-} \ ,
     \label{I.1a}
\end{equation}
where $\omega_{\rm A^{\sigma}}$ are the occupation numbers and
$E_{A^{\sigma}}$ are the energies for the integer charge species
$\sigma =$ +, 0, and $-$.  The $\omega_{A^{\sigma}}$ depend on $q$ and
are equal to or greater than zero.  (See Refs.~\onlinecite{ParrYang},
\onlinecite{ToddM}, and \onlinecite{PerdewNATO} for the detailed
expressions.)  MM show that the GC energy as expressed by
Eq.~(\ref{I.1a}) minimizes to integer charges, and never to fractional
charges.  When the two atoms are well separated and only weakly
interacting, the covalent and ionic resonance states can be regarded as
being close to eigenfunctions and the mixing terms between the states
can be ignored.  The nonlinearities in the occupation numbers as
functions of charge introduced through temperature at physically
reasonable temperatures are not sufficient to produce states with
fractional charges.  Thus charge transfer as represented by an ensemble
of weakly interacting integer-charge states is too simplistic to
describe fractionally charged states.

MM also examine a 3-state valence bond (VB) approach usqing the same
resonance states as in the GC approach.  This is equivalent to a 0 K
ensemble where the states are allowed to interact.  The VB approach is
able to represent charge transfer processes with fractional charges.
However, the representation of the energy depends on resonance matrix
elements.  The issue becomes one of defining sets of coefficients for
the state wavefunctions that allow one to recover quadratic and
GC-like energy expressions.  Using Coulson
charges\cite{Coulson,Murrell,McWeeny} as an approximation and a
maximum entropy valence bond approach (MEVB), MM derive energy
expressions as functions of $q$ with the forms of Eqs.~(\ref{I.1}) and
(\ref{I.1a}).  In the MEVB approach, decomposition into atomic
contributions is accomplished through an examination of the various
matrix elements appearing in the total energy expression.  However, MM
were unable to find a general expression linking the GC and classical
electrostatic forms.

The three main difficulties in deriving a charge dependent potential
are the same ones facing MM and others.  First is to define the charge,
second is to evaluate or eliminate the resonance energy matrix
elements, and third is to define the energy for a particular value of
$q$.  First, to define $q$, we invoke a
density-functional-theory\cite{HK,KS,Levy} (DFT) motivated
atom-in-molecule\cite{hir,Moffit,Bader} pseudo-atom concept within the context
of the EVB approach.  We assume the availability of a practical density
decomposition strategy\cite{PDL78,Guse,LiParr,SCAD,hir,Bader,ATV02} to
define the pseudo-atom densities.  The charge is defined as an average
over the difference between two pseudo-atom densities.  No restriction
to Coulson\cite{Coulson} or other definitions of
charge\cite{Lowdin,Mulliken} is necessary.  The dependence of the EVB
wavefunction on $q$ is then deduced, which in turn yields an energy for
arbitrary $q$, not just the optimum $q$.  Second, to evaluate resonance
energy matrix elements, we retain the VB approach of MM, but use an
empirical valence bond (EVB)\cite{DC,Warshel,Try} strategy, rather than
explicitly evaluating a model Hamiltonian or using the MEVB averaging
procedure of MM. Specifically, a reference energy is separated out for
a fixed value of $q$.  To define the energy for an atom in the
molecule, we require consistency between the EVB wavefunction energy
and DFT energy for each atom.  In so doing we are able to cast the
energy of each atom into a form suitable for constructing empirical
potential energy surfaces that could be used in simulations of larger
systems.  Third, the typical situation
is that there are many wavefuctions and many electron densities that
are consistent with a particular $q$.  To define a unique energy from
among the possible choices, we adhere to constrained search density
functional theory (CS-DFT)\cite{Levy,SMV} rather than appealing to a
maximum entropy principle.

To the best of our knowledge, the EVB method has not
previously been combined with an atom-in-molecule
approach.\cite{Geerlings} von Szentp\'{a}ly and coworkers define atomic
charges, but their definition is implicitly limited to a quadratic
dependence.\cite{vSNG1,vSNG2,vSNG6} Grochowski and coworkers\cite{Try}
use EVB parameterized by {\it ab initio} calculations for key molecular
fragments and for defining the charges.  The charges are defined with a
phenomenological spatial dependence that does maintain correct
dissociation to neutrals.  Furthermore, the charge dependence of the
final potential is purely electrostatic.

The inconsistency between the GC and classical electrostatic forms is
explained in Cios{\l}owski and Stefanov\cite{CioStef} using a different
definition of $q$, which is based on the total system wavefunction.
The charge is expressed as a perturbation on the molecular Hamiltonian
$H$.  Computed atom-in-molecule equilibrium charges are used in
subsequent ``charge-constrained'' calculations to study the energy and
electronegativity dependences about the ground-state atom-in-molecule
charges.  Nalewajski\cite{Nal-KS} provides a simplified rendition of
Cios{\l}owski and Stefanov, but only for orthogonal resonance states.
While physically correct, these efforts have not been cast in the form
of general purpose potential energy surfaces.  In particular, neither
attempts to resolve the resonance energy issue in a way that is
tractable for large-scale simulations.

Here we derive a charge-dependent empirical potential which faithfully
represents charge transfer as a function of separation between atom A
and an entity B. In the simplest case, entity B is another atom.  More
generally, B represents a collective embedding environment or
reservior.\cite{PPLB,NalParr} The charge may be fractional.  Our
primary interest is in deriving a general functional form with correct
physical and chemical behavior at all interaction strengths rather than
providing an exact treatment of particular terms or systems.  The three
difficulties just outlined are addressed.  We then analyze these
potentials in special limiting cases and in the light of the results of
Cios{\l}owski and Stefanov\cite{CioStef} and Nalewajski.\cite{Nal-KS}
Finally, we derive general models for both pair potentials and
atom-in-molecule energies.

\section{The Empirical Valence Bond Representation}
The EVB method is a much more general technique than described here.
Reviews of EVB are available from Warshel {\it et.~al}.\cite{Warshel}
Here we confine our discussion to a 2-state model of the AB molecule.
There are fixed covalent and ionic resonance states represented by
wavefunctions ${\psi_c}$ and ${\psi_i}$,
respectively.\cite{Warshel,Try,MullikenDi,Jeremy} No assumption is
made about the quality of these wavefunctions.  However, ${\psi_c}$
retains $N_{\rm A^0}$ electrons on atom A and $N_{\rm B^0}$ electrons
on atom B. Atoms A and B are neutral in the covalent resonance state.
Consequently $N_{\rm A^0}$ and $N_{\rm B^0}$ are equal to their
respective nuclear charges, $Z_{\rm A}$ and $Z_{\rm B}$.  Similarly,
${\psi_i}$ retains $N_{\rm A^0}-1$ electrons on atom A and $N_{\rm
B^0}+1$ electrons on atom B. In each resonance state, the total number
of electrons is $N$.  ${\psi_c}$ and ${\psi_i}$ are assumed to be
normalized to unity.  The wavefunction of the system $\psi$ is the
combination\cite{Weinbaum,Coulson,CoulFisch,Murrell,McWeeny,BarbShukla}
\begin{equation}
     \psi = c \, (\psi_c + \gamma \, \psi_i) ,  \label{II.2}
\end{equation}
where $\gamma$ determines the relative ionic character.  By
normalization,
\begin{equation}
     1/c^2 = 1 + 2 \, \gamma \, S_{ci} + \gamma^2 \ ,  \label{II.3}
\end{equation}
where
\begin{equation}
     S_{ci} = \langle\psi_c|\psi_i\rangle \ .  \label{II.4}
\end{equation}
For the AB molecule with Hamiltonian $H$, the mixed-state energy takes
the form
\begin{eqnarray}
     E &=& c^2 \, (H_{cc} + 2 \, \gamma \, H_{ci} + \gamma^2
     H_{ii}) \ , \label{II.5} \\
     &=& \frac{H_{cc} + 2 \, \gamma \, H_{ci} + \gamma^2 H_{ii}}{1 +
     2 \, \gamma \, S_{ci} + \gamma^2} \ , \label{II.6}
\end{eqnarray}
where $H_{cc}$, $H_{ii}$ and $H_{ci}$ are the associated energy matrix
elements,
\begin{eqnarray}
     H_{cc} &=& \langle\psi_c|H|\psi_c\rangle \ ,  \label{II.6a} \\
     H_{ii} &=& \langle\psi_i|H|\psi_i\rangle \ ,  \label{II.6b}
\end{eqnarray}
and
\begin{eqnarray}
     H_{ci} = \langle\psi_c|H|\psi_i\rangle \ . \label{II.6c}
\end{eqnarray}

Minimizing $E$ with respect to $\gamma$ gives the optimized
values,
\begin{equation}
     \gamma_{\rm opt} = \frac{1 \pm \sqrt{1 + \epsilon_{cc} \,
     \epsilon_{ii}}}{\epsilon_{ii}} \ , \label{II.7}
\end{equation}
where $\epsilon_{cc} = 2 \, (H_{ci}-S_{ci} \, H_{cc})/(H_{ii}-H_{cc})$
and $\epsilon_{ii} = 2 \, (H_{ci}-S_{ci} \, H_{ii})/(H_{ii}-H_{cc})$.
The $\pm$ signs in Eq.~(\ref{II.7}) correspond to ground (gs) and
excited (xs) states, whose energies are designated as $E_{\rm gs}$ and
$E_{\rm xs}$, respectively.  From Eq.~(\ref{II.7}), one can see that
the off-diagonal matrix elements, $H_{ci}$ and $S_{ci}$, control charge
transfer in EVB. When $S_{ci} = 0$, $\epsilon_{cc} = \epsilon_{ii} =
\epsilon = 2 \, H_{ci}/(H_{ii}-H_{cc})$.  Depending on the root,
$\gamma_{\rm opt}$ ($-$ root) or $1/\gamma_{\rm opt}$ (+ root) varies
from $-$1 to +1.  Note that as $H_{ii}-H_{cc}$ goes to zero, that is,
as the covalent and ionic curves cross, charge transfer or the
Coulson-Fischer transition\cite{CoulFisch} becomes more abrupt.  If
$\psi_c$ and $\psi_i$ are nondegenerate eigenfunctions of $H$, then
$H_{ci} = S_{ci} = 0$.  There is either no charge transfer or there is
complete charge transfer.  In either case, the states do not mix.

The coefficient $\gamma$ governs the ionic contribution to $\psi$.  As
$\gamma$ increases, $\psi$ migrates toward more complete charge
transfer.  Even neglecting overlap,\cite{DC,Warshel} the ionic strength
$\gamma$ interpolates between covalent and ionic states in a physically
reasonable way.\cite{CoulFisch} The state-to-state interpolating
behavior of $\gamma$ in Eq.~(\ref{II.7}) is the essential behavior that
we wish to emulate in developing a more broadly applicable charge
dependent potential.

Later in the paper, two other relationships will become useful, which
we provide here.  First, in the spirit of EVB,\cite{Warshel} one always
wants to know the resonance energy in the terms of the ground-state
$E_{\rm gs}$, $S_{ci}$, $H_{cc}$, and $H_{ii}$.  That is,
\begin{eqnarray}
     H_{ci} = E_{\tau} \, S_{ci} \pm \sqrt{(H_{cc} - E_{\tau}) (H_{ii}
     - E_{\tau})} \ ,  \label{II.8}
\end{eqnarray}
where $\tau$ is either ``gs'' or ``xs''.  The second relationship
expresses $\gamma_{\rm opt}$ as a function of these same parameters.
This is done by substituting Eq.~(\ref{II.8}) into the expressions for
$\epsilon_{cc}$ and $\epsilon_{ii}$.  The general expression is
\begin{eqnarray}
     && \negthickspace \negthickspace \negthickspace \negthickspace
     \epsilon_{\sigma\sigma} = \nonumber \\
     &2& \Bigg[ \frac{(E_{\tau} - H_{\sigma\sigma}) \,
     S_{ci} \pm \sqrt{(H_{cc} - E_{\tau})
     (H_{ii} - E_{\tau})}}{H_{ii}-H_{cc}} \Bigg] \ ,
     \nonumber \\
     \label{II.9}
\end{eqnarray}
where $\sigma\sigma$ is either ``$cc$'' or ``$ii$'' and either the
ground or excited state is selected.

\section{Charge in the 2-State EVB Model}
We begin with the question of how to define charge in our model.  There
is no unique definition of the charge on an atom in a molecule.  To
assign charges to individual atoms, we assume that it is possible to
decompose the total electron density into pseudo-atom densities.
However, we do not need to specify a particular decomposition procedure
at this time.  We only need to know that some procedure is
available.\cite{PDL78,Guse,LiParr,SCAD,hir,Bader,ATV02}

\subsection{Definition of Charge}
To define the charge, it is convenient to use $N$-electron density
matrices and 1-electron densities.  In terms of density matrix
language, the state of AB is represented as
\begin{equation}
     \Gamma ({\bf r}_N', {\bf r}_N) = \psi({\bf r}_N') \, \psi({\bf
     r}_N) \ , \label{IV.1}
\end{equation}
where ${\bf r}_N$ are the 3$N$ dimensional spatial electronic
coordinates for the full AB system and $\psi$ is given in
Eq.~(\ref{II.2}).  Spin is ignored at this point and, for simplicity,
the matrix elements are assumed to be real.  Eq.~(\ref{IV.1})
corresponds to the pure-state representation of the density matrix for
the 2-state model.  We can expand Eq.~(\ref{IV.1}) in terms of the
covalent and ionic states, resulting in the relationship
\begin{equation}
     \Gamma = \alpha_{cc} \, \Gamma_{cc} + 2 \,\alpha_{ci} \,
     \Gamma_{ci} + \alpha_{ii} \, \Gamma_{ii} \ , \label{IV.2}
\end{equation}
where
\begin{eqnarray}
     \Gamma_{cc} &=& \psi_c^2 \ , \nonumber \\
     \Gamma_{ci} &=& \psi_c \, \psi_i \ , \nonumber \\
     \Gamma_{ii} &=& \psi_i^2 \ , \nonumber \\
     \alpha_{cc} &=& 1/(1 + 2 \, \gamma \, S_{ci} + \gamma^2) \ ,
     \nonumber \\
     \alpha_{ci} &=& \gamma/(1 + 2 \, \gamma \, S_{ci} + \gamma^2) \ ,
     \nonumber
\end{eqnarray}
and
\begin{eqnarray}
     \alpha_{ii} = \gamma^2/(1 + 2 \, \gamma \, S_{ci} + \gamma^2) \ .
     \nonumber
\end{eqnarray}
The total electronic energy expression analogous to Eq.~(\ref{IV.2})
is
\begin{eqnarray}
     E = \alpha_{cc} \, H_{cc} + 2 \, \alpha_{ci} \, H_{ci} +
     \alpha_{ii} \, H_{ii} \ .  \label{IV.3}
\end{eqnarray}
Eq.~(\ref{IV.3}) corresponds to a non-diagonal representation of the
energy.  Its advantage is that the component contributions of the
essential states that are thought to represent the physical system are
delineated.  A diagonal representation of $\Gamma$ when state mixing
is important can be devised by diagonalizing $H$.  For the purposes of
extending the present treatment to finite temperature
ensembles,\cite{PPLB,PerdewNATO,ToddM,CedParr} this
particular diagonalization of $\Gamma$ would be useful.

The $N$-electron densities of interest are $\rho$, $\rho_{cc}$,
$\rho_{ci}$, and $\rho_{ii}$.  They bear the usual relationships to
% the respective $N$-electron density matrices, $\Gamma$, $\Gamma_{cc}$,
% $\Gamma_{ci}$ and $\Gamma_{ii}$:
the respective $N$-electron density matrices, $\Gamma_{\sigma \tau}$:
\begin{eqnarray}
     \rho_{\sigma \tau}({\bf r}) = N \int d{\bf r}_{N-1}\,
     \Gamma_{\sigma \tau} ({\bf r}_N, {\bf r}_N) \ ,
\end{eqnarray}
where ${\bf r}_{N-1}$ are the 3($N$-1) dimensional spatial electronic
coordinates and ${\sigma \tau}$ is either ``$cc$'', ``$ci$'',
``$ii$'', or no subscript.  $\rho$, $\rho_{cc}$, and $\rho_{ii}$ are
normalized to $N$.  The relationship of central interest is
\begin{eqnarray}
     \rho &=& \alpha_{cc} \, \rho_{cc} + 2 \, \alpha_{ci} \, \rho_{ci} +
     \alpha_{ii} \, \rho_{ii} \nonumber \\
     &=& \frac{\rho_{cc} + 2 \, \gamma \, \rho_{ci} + \gamma^2
     \rho_{ii}}{1 + 2 \, \gamma \, S_{ci} + \gamma^2}\ .
     \label{III.1}
\end{eqnarray}
Recall the assumption that the covalent and ionic state
wavefunctions are given and fixed.  Thus the total
density $\rho$ is determined solely by the value of $\gamma$.

The energies of $\rho_{cc}$ and $\rho_{ii}$ are well-defined in a
conventional DFT sense.\cite{HK,Levy} However, the ``interference
density''\cite{RuedRMP} $\rho_{ci}$ does not have a well-defined energy
in DFT. Nevertheless, its energy may be inferred from the energies of
$\rho$, $\rho_{cc}$, and $\rho_{ii}$, as we will show below.

Next we assume that all of the $\rho_{\sigma \tau}$ can be decomposed
into corresponding pseudo-atom densities, $\rho_{\sigma \tau, \rm A}^*$
and $\rho_{\sigma \tau, \rm B}^*$.\cite{RuedRMP} The pseudo-atom
densities $\rho_{\rm A}^*$ and $\rho_{\rm B}^*$ integrate to
non-integer values, $N_{\rm A}^*$ and $N_{\rm B}^*$, whereas
$\rho_{cc,\rm A}^*$, $\rho_{ii,\rm A}^*$, $\rho_{cc,\rm B}^*$, and
$\rho_{ii,\rm B}^*$ are constrained to integrate to integer numbers of
electrons.  We use asterisks throughout to indicate atom-in-molecule
quantities.

With these definitions in place, we define $q$ from either pair of
total and covalent pseudo-atom densities, which others have sometimes
referred to as pseudo-atom distortion densities.\cite{hir,Bader} We
choose atom A:
\begin{eqnarray}
     q &=& \int d{\bf r}\, (\rho_{cc,\rm A}^*({\bf r}) - \rho_{\rm
     A}^*({\bf r})) \ \label{III.2} \\
     &=& N_{\rm A} - N_{\rm A}^* \ .
\label{I.14}
\end{eqnarray}
The density decompositions must be constrained to yield the correct
number of electrons prescribed by
Eq.~(\ref{I.14}).\cite{NalParr2,NalParr3,NalLos} Finally, we note that
the present definition of $q$ falls into Truhlar's Class II
category.\cite{DGT0,DGT1}
% Note that the integration in Eq.~(\ref{III.2}) is over all
% space,\cite{hir} whereas Bader defines pseudo-atoms in finite regions
% of space.\cite{Bader}

Now we want to eliminate $\gamma$ in favor of $q$.  Assuming that a
component definition of $\rho_{\rm A}^*({\bf r})$ based on
Eq.~(\ref{III.1}) is possible,
\begin{equation}
     q = \frac{\gamma^2 - 2 \, \gamma \, \delta N_{ci,\rm A}^*}{1 + 2
     \, \gamma \, S_{ci} + \gamma^2} \ , \label{III.3}
\end{equation}
where
\begin{equation}
     \delta N_{ci,\rm A}^* \equiv \int d{\bf r} \, (\rho_{ci,\rm A}^*({\bf
     r}) - (N_{A^0}/N) \, \rho_{ci}({\bf r})) \ ,  \label{III.4}
\end{equation}
$N_{A^0}$ is the number of electrons on atom A when it is in a neutral
state, and the relationship $N \, S_{ci} = \int d{\bf r} \,
\rho_{ci}({\bf r})$ has been used.  The quantity $\delta N_{ci,\rm
A}^*$ is determined by the difference between $\rho_{ci,\rm A}^*$ and
the atom A component of the decomposition of $\rho_{ci}$ with locally
unbiased, electron-number decomposition.\cite{NalParr3} Clearly,
different density decomposition strategies will yield somewhat
different values of $\delta N_{ci,\rm A}^*$.  One possibility for
determining $\delta N_{ci,\rm A}^*$ is to require consistency with the
ground-state value of $\gamma$.  For instance, if $q = 0$ is optimum
as for a dimer, $\gamma$ equals either 0 or $2 \, \delta N_{ci,\rm
A}^*$.  For Eq.~(\ref{III.3}) to be applied successfully, $2 \, \delta
N_{ci,\rm A}^*$ would have to correspond to a lower energy state than
$\gamma = 0$ and the value of $\gamma$ would have to be determined from
a separate calculation, such as represented by Eq.~(\ref{II.7}).  In
such an approach, one would be effectively modeling $\delta N_{ci,\rm
A}^*$ via a correspondence with the Coulson-Fischer
transition.\cite{CoulFisch} Alternatively, by analogy with $\rho$, we
assume that single-particle determinants (e.g.~Kohn-Sham
determinants\cite{KS}) can be calculated for the $\rho_{\sigma \sigma,
\rm A}^*$.\cite{ZMP,WangParr,SCAD,PWAmax} In a Kohn-Sham based
approach, one would be effectively estimating $\rho_{ci,\rm A}^*$ from
the overlaps of these determinants.  Additional ambiguity in $\delta
N_{ci,\rm A}^*$ arises from the choices for $\psi_c$ and $\psi_i$.
These ambiguities lie behind the designation of the present approach as
an empirical one.  However, these ambiguities can be mitigated by using
a reference state as discussed in the next Subsection.  To that end, it
is useful to invert Eq.~(\ref{III.3}) so that $\delta N_{ci,\rm A}^*$
becomes a function of $q$ and $\gamma$.  That relationship is
\begin{equation}
     \delta N_{ci,\rm A}^* = \frac{(1-q) \, \gamma^2 - 2 \, S_{ci} \,
     \gamma -q}{2 \, \gamma} \ , \label{III.10}
\end{equation}
It should be understood that, in the limit that $\gamma \rightarrow
0$, $\delta N_{ci,\rm A}^* \rightarrow 0$ also.

\subsection{Constructing Pair Potentials}
To construct a potential energy surface for AB, one option is to use
Eq.~(\ref{III.3}) to model the dependence of the charge on separation
for some reference state.  Some empirical potentials such as EVB and
the modified embedded atom method (MEAM)\cite{MEAM1,MEAM2,MEAM3}
utilize reference states as a model calibration method.  The methods of
McDonald and coworkers,\cite{Imac} McCammon, Grochowski, and
coworkers,\cite{Try} and Broughton and Mehl\cite{Jeremy} effectively
make $q$ bond-length dependent.  Eq.~(\ref{III.3}) provides a basis in
EVB theory for their phenomenological charge transfer switching
functions.

A more attractive option is to solve for $\gamma = \gamma(q)$,
\begin{equation}
     \gamma = \frac{(\delta N_{ci,\rm A}^* + q \, S_{ci}) \pm
     \sqrt{(\delta N_{ci,\rm A}^* + q \, S_{ci})^2 + q\, (1-q)}}{1-q} \
     , \label{III.5}
\end{equation}
The coefficient $\gamma$ determines the strength of the contribution
of $\psi_i$ to $\psi$.  Eq.~(\ref{III.5}) states how the charge
governs that strength.  This expression is consistent with the results
of Cios{\l}owski and Stefanov,\cite{CioStef} which are derived from a
perturbative technique.  As noted previously, even if $q = 0$,
$\gamma$ equals either 0 or $2 \, \delta N_{ci,\rm A}^*$.  This is
because the EVB model describes state mixing even when there is no
charge transfer.  For example, in the $\rm H_2$ molecule, the covalent
and ionic wavefunctions mix at all finite separations, but the ground
state never involves charge transfer.  Significantly, this formula
also describes deviations from the ground-state charge.

Eq.~(\ref{III.5}) can be substituted into Eq.~(\ref{II.6}), and the
variational procedure repeated.  The result is the same as solving for
$q$ in terms of the resonance and overlap matrix elements obtained by
equating Eq.~(\ref{III.5}) and (\ref{II.7}).

The EVB strategy is to use experimental information to eliminate the
resonance energy.\cite{DC,Warshel} Here the analogous procedure is to
choose a particular value of $q = q_0$ and solve for $H_{ci}$ in terms
of $E(q_0)$ and the diagonal matrix elements for each $R$.
(The $R$ dependence is suppressed.)  The result is
\begin{eqnarray}
     H_{ci} = \frac{E(q_0) - \alpha_{cc}(q_0) \, H_{cc} -
     \alpha_{ii}(q_0)\, H_{ii}}{2 \, \alpha_{ci}(q_0)} \ .
     \label{III.5a}
\end{eqnarray}
Substituting Eq.~(\ref{III.5a}) into Eq.~(\ref{IV.3}), the total energy
for arbitrary $q > 0$ has the form
\begin{widetext}
\begin{equation}
     E(q) = (\alpha_{ci}(q)/\alpha_{ci}(q_0)) \, E(q_0) +
     (\alpha_{cc}(q) - \alpha_{cc}(q_0) \,
     \alpha_{ci}(q)/\alpha_{ci}(q_0)) \, H_{cc} + (\alpha_{ii}(q) -
     \alpha_{ii}(q_0) \, \alpha_{ci}(q)/\alpha_{ci}(q_0)) \, H_{ii} \ .
  \label{III.5b}
\end{equation}
\end{widetext}
This form might be used in lieu of classical electrostatic potentials
like Eq.~(\ref{I.1}) that have been in common use.  It has the
structure of an ensemble average,\cite{PPLB} but the coefficients of
$H_{cc}$ and $H_{ii}$ are not necessarily positive semidefinite.
Eq.~(\ref{III.5b}) is constructed to possess the proper changes in
atomic charges in the limit of molecular dissociation.  The
construction of a proper ensemble representation is discussed in
Section 5.A.

The foundations for the quadratic dependence of the energy on charge,
such as Eq.~(\ref{I.1}), must stem from Eqs.~(\ref{III.5}) and
(\ref{II.6}).  However, even this simplest example of charge transfer
has a considerably more complex dependence than quadratic.  That
dependence is clearly carried through the overlap contributions.

Once $q_0$ has been chosen, the procedure for determining a point on
the potential energy surface for arbitrary values of $R$ and $q$ is as
follows.  From some other source(s) of information, one must have
available five reference values: $E(q_0)$, its associated ionicity
$\gamma_0$, $H_{cc}$, $H_{ii}$, and $S_{ci}$.  For a chosen $R$, one
first evaluates
\begin{itemize}
     \item [(a)] $\alpha_{cc}$, etc. from Eq.~(\ref{IV.2});

     \item [(b)] $H_{ci}$ from Eq.~(\ref{III.5a}); and

     \item [(c)] $\delta N_{ci,\rm A}^*$ from Eq.~(\ref{III.10}).
     \begin{flushleft}
	Then, for each $q$ of interest, one evaluates
     \end{flushleft}
     \item [(d)] $\gamma(q)$ from Eq.~(\ref{III.5}) using values from
     Steps (a), (b), and (c); and

     \item [(e)] $E(q)$ from Eq.~(\ref{II.5}).
\end{itemize}
The procedure is repeated for each value of $R$ of interest.

An important question is what to choose for $q_0$.  One possible choice
for $q_0$ is the optimum value $q_{\rm opt}$.  However, $q_{\rm opt}$
depends on $H_{ci}$ through Eqs.~(\ref{II.7}) and (\ref{III.3}).  This
variant is equivalent to using the ground-state wavefunction as one of
the basis functions in the original formulation of the problem.
Eq.~(\ref{III.5b}) then characterizes deviations of the energy from the
ground-state energy, $E(q_{\rm opt})$, as a function of $q$.  Note
that by choosing $E(q_0)$ to be consistent with $E(q_{\rm opt})$, this
variant and Eq.~(\ref{III.5b}) will be identical.

When $q_0 = q_{\rm opt}$, the procedure for determining a point on the
potential energy surface is substantially the same as the first.  Again
five reference values are needed, except that knowledge of $q_{\rm
opt}$ replaces knowledge of $\gamma_0$.  Relative to the first
procedure, Steps (a) and (b) become to evaluate
\begin{itemize}
     \item [(a)] $H_{ci}$ from Eq.~(\ref{II.8}); and

     \item [(b)] $\gamma_{\rm opt}(q_{\rm opt})$ from Eq.~(\ref{II.7}).
\end{itemize}
All of the other steps remain the same.

One can extend this model to $q$ between $-1$ and 0.  The entire
procedure with atom A assumed to be anionic in $\psi_i$ is repeated.
In Eq.~(\ref{I.14}), for $q < 0$, one simply replaces $q$ with $-q$ and
adjusts the partitioning of $\rho_{ci}$ so that atom A is anionic.
Requiring continuity in the energy at $q = 0$ dictates that $-\delta
N_{ci,\rm A}^{*} \rightarrow \gamma(q)$ as $q \rightarrow 0^-$.
Eq.~(\ref{III.5b}) remains the same structurally.  Equivalently, one
could replace the subscript ``A'' with the subscript ``B'' everywhere
in the procedure.  The potential over the entire range of $q$ is then
represented in a piecewise fashion.  While not as rigorous as a 3-state
model, the present treatment does retain substantially greater
simplicity.

Another possible extension is to apply the above procedure when the two
resonance states both correspond to charged species.  For CaO, for
example, the effective $q$ on Ca near equilibrium would be almost +2.
As the CaO bond is stretched, $q$ would decrease until it passed
through a region of undetermined length where it would range between +1
and 0.  Then there would be one form of Eq.~(\ref{III.5b}) covering the
range in $q$ between 0 and +1 and a second form covering the range
between +1 and +2 range.

A final extension of this model would allow the entity B correspond to
a more general environment than just one other atom.  Most of the model
presented here does not explicitly invoke the specific properties of a
diatomic model.  However, this extension requires separate
considerations not pursued here.

\section{Definition of Pseudo-Atom Energies}
We now show how to define pseudo-atom energies for the 2-state model of
the previous section.  We do this by requiring consistency between the
energies based on the density decompositions that were assumed in the
previous Section and the energies that would result from the
corresponding wavefunction expressions.  As with $\rho$, we assume that
there is a decomposition $\Gamma$ into pseudo-atom density matrices.
In conformance with Rychlewski and Parr,\cite{RychParr} the
decomposition applies to $\Gamma$ rather than to $H$.  Thus, for
\begin{eqnarray}
     \Gamma &=& \Gamma_{\rm A}^* + \Gamma_{\rm B}^* \ , \label{IV.4}
\end{eqnarray}
the total energy decomposes into
\begin{eqnarray}
     E = \langle H,\Gamma_{\rm A}^*\rangle + \langle H,\Gamma_{\rm
     B}^*\rangle = E_{\rm A}^* + E_{\rm B}^* \ .
     \label{IV.4a}
\end{eqnarray}
Analogous expressions are assumed to exist for each of the covalent and
ionic contributions to the pseudo-atom energies.

The pseudo-atom densities $\rho_{\rm A}^*$ and $\rho_{\rm B}^*$
correspond to the pseudo-atom density matrices $\Gamma_{\rm A}^*$ and
$\Gamma_{\rm B}^*$.  In the following relations, the expressions for
atoms A and B are analogous.  Only the expressions for atom A will be
given.  We want the energies of each pseudo-atom defined via density
matrices to be equal to the energies defined via densities.
Consequently, we require that
\begin{equation}
     E_{\rm A}^* = E_{A}[\rho_{\rm A}^*] \ .  \label{IV.5a}
\end{equation}
This identification places a new constraint on $\Gamma_{\rm A}^*$.
Rigorously speaking, it should be optimal in the sense of Levy
CS-DFT.\cite{Levy,SMV}  The constraint is that
$\Gamma_{\rm A}^*$ should yield the lowest energy for all
other ensemble density matrices which integrate to $\rho_{\rm A}^*$:
\begin{equation}
     E_{A}^{\rm CS-DFT}[\rho_{\rm A}^*] \equiv \langle H,\Gamma_{\rm
     A}^*\rangle = \min_{\Gamma_{\rm A} \rightarrow \rho_{\rm A}^*}
     \langle H,\Gamma_{\rm A}\rangle \ .  \label{IV.5c}
\end{equation}

In terms of valence bond resonance states
\begin{eqnarray}
     E_{\rm A}^* = \alpha_{cc} \, H_{cc,\rm A}^* + 2 \, \alpha_{ci} \,
     H_{ci,\rm A}^* + \alpha_{ii} \, H_{ii,\rm A}^* \ .  \label{IV.6}
\end{eqnarray}
Similar to $E_{\rm A}^*$, the pure-state terms in Eqs.~(\ref{IV.6}),
$H_{cc,\rm A}^*$ and $H_{ii,\rm A}^*$, can be represented in
conventional DFT language.  The resonance energy cannot.
Eq.~(\ref{IV.6}) represents pseudo-atom energies for any $q$.  To
evaluate the resonance energy, we again follow the EVB strategy of
determining them from some particular value of $q = q_0$.  The result
is
\begin{eqnarray}
     H_{ci,\rm A}^* = \frac{E_{\rm A}^*(q_0) - \alpha_{cc}(q_0) \,
     H_{cc,\rm A}^* - \alpha_{ii}(q_0)\, H_{ii,\rm A}^*}{2 \,
     \alpha_{ci}(q_0)} \ .  \label{IV.7a}
\end{eqnarray}
If $q_0$ is set equal to the optimum $q$ for each value of atomic
separation, then $E(q_0) = E_{\rm A}^*(q_0) + E_{\rm B}^*(q_0)$ will
correspond to the experimental potential energy for AB. Note that all
of the energies in traditional EVB are 0 K values.  Some variations of
EVB incorporate temperature-dependent solvent effects.\cite{Warshel} We
do not include these variations here.  Finite temperatures are not
required to establish the model.  On the other hand, there is nothing
here that precludes extending the analysis to a finite temperature
ensemble.\cite{PPLB,NalParr,Nal-KS,ToddM,CedParr}

Substituting Eq.~(\ref{IV.7a}) into Eq.~(\ref{IV.6}), an expression
for the pseudo-atom energy is achieved.  For any $q$,
\begin{eqnarray}
     E_{\rm A}^*(q) &=& (\alpha_{ci}(q)/\alpha_{ci}(q_0)) \, E_{\rm
     A}^*(q_0) \nonumber \\ &+& (\alpha_{cc}(q) - \alpha_{cc}(q_0) \,
     \alpha_{ci}(q)/\alpha_{ci}(q_0)) \, H_{cc,\rm A}^* \nonumber \\
     &+& (\alpha_{ii}(q) - \alpha_{ii}(q_0) \,
     \alpha_{ci}(q)/\alpha_{ci}(q_0)) \, H_{ii,\rm A}^* \ .  \nonumber
     \\ \label{IV.8a}
\end{eqnarray}
Eq.~(\ref{IV.8a}) has the same structure as Eqs.~(\ref{III.5b}), but
refers to an individual atom.  All of the quantities on the right-hand
sides of these two equations can be deduced from experiment and/or
decomposition calculations on resonance states and overlaps.

Using the present formulations in simulations of larger systems
naturally invokes consideration of chemical potential equalization.
One can obtain a statement of chemical potential equalization from
Eq.~(\ref{IV.8a}).  The total energy is given in Eq.~(\ref{IV.4a}).  If
$E$ is minimized with respect to $q$, then small deviations from
$q_{\rm opt}$ will not change the total energy to first order:
\begin{eqnarray}
0 &=& \frac{dE(q)}{dq} \Big|_{q=q_{\rm opt}} \nonumber \\
&=& \frac{dE_{\rm A}^*(q)}{dq} \Big|_{q=q_{\rm opt}}  +
\frac{dE_{\rm B}^*(-q)}{dq} \Big|_{q=q_{\rm opt}}  \ , \label{IV.9}
\end{eqnarray}
or
\begin{eqnarray}
\frac{dE_{\rm A}^*(q)}{dq} \Big|_{q=q_{\rm opt}} = \frac{dE_{\rm
B}^*(-q)}{d(-q)} \Big|_{q=q_{\rm opt}} \equiv - \mu(q) \ ,
\label{IV.10}
\end{eqnarray}
where $\mu$ is the chemical potential.  Charge balance requires that
the charges on A and B be exactly opposite.  The minus sign in front
of $\mu$ comes from the fact that $q$ is related to the negative of
the change in the number of electrons on A.

\section{Discussion and Examples}
Here we introduce approximations consistent with the dissociation
limits of AB, present a general definition of the energy for a given
charge, and discuss $\rm {H_2}$, HF, and LiH as examples.

\subsection{Neglect of Differential Overlap Model}
It is insightful to introduce a concept of neglect of differential
overlap between resonance states, $\rho_{ci} = 0$, which we will refer
to as NDOL. This leads to simple analytical expressions whose behavior
can be examined in detail.  NDOL is to be distinguished from zero
differential overlap (ZDO),\cite{Coulson} which refers to overlap
between orbitals on different atomic centers.  In the NDOL
approximation, $\rho_{ci} = 0$.\cite{ParksParr} Under special
conditions, ZDO implies NDOL.

In the NDOL approximation, $\delta N_{ci,\rm A}^{*}$ and $S_{ci}$ are
zero.  From Eq.~(\ref{III.5}), the dependence of $\gamma$ on $q$
becomes
\begin{equation}
     \gamma(q)=\pm \sqrt{q/(1-q)}
     \ . \label{III.6.1}
\end{equation}
 From Eq.~(\ref{II.6}), the energy dependence on $q$ becomes
\begin{eqnarray}
     E^{\rm NDOL}(q) &=& H_{cc} - 2 \sqrt{q \, (1-q)} |H_{ci}|
     \nonumber \\
     &+& q\, ({\rm IP}_{\rm A}^* - {\rm EA}_{\rm B}^*) \ .
     \label{III.6}
\end{eqnarray}
The $-|H_{ci}|$ construct ensures that $E^{\rm NDOL}$ corresponds to
the ground state.  Here we have used the fact that, by definition of
the ionization potential IP and electron affinity EA, $H_{ii} - H_{cc}
= {\rm IP}_{\rm A}^* - {\rm EA}_{\rm B}^*$.  These atom-in-molecule
quantities include some electrostatic contributions.  This quantity is
also called a ``bond hardness''.\cite{CioStef}

Thus, Eq.~(\ref{III.6.1}) is consistent with previous energy
expressions obtained with Coulson
charges\cite{Coulson,Murrell,McWeeny,ToddM} and with the 2-state model
of Nalewajski\cite{Nal-KS}.  In Coulson,\cite{Coulson} Murrell {\it
et.~al},\cite{Murrell} and McWeeny\cite{McWeeny} a fraction of ionic
character is defined instead of a charge.  That fraction is identical
to $\gamma^2/(1+\gamma^2)$ which is $q$ in the NDOL approximation.
Clearly, at large $R$, Eq.~(\ref{II.5}) in combination with
Eq.~(\ref{III.5}) approaches Eq.~(\ref{III.6}), Eq.~(\ref{III.6})
becomes linear in $q$, and $H_{ii}$, $H_{cc}$, and $H_{ci}$ approach
asymptotic values (in $R$) analogous to MM's Eq.~(2.40).\cite{ToddM}
The linear behavior in the asymptotic regime is consistent with the
conclusions of PPLB. Even more importantly, Eq.~(\ref{III.6}) is
expressly non-analytical ({\it i.e.}, it cannot be expanded in a Taylor
series) about $q = 0$ and $q = 1$, whereas Eq.~(\ref{II.5}) in
combination with Eq.~(\ref{III.5}) is analytical at both points away
from the NDOL limit ($\delta N_{ci,\rm A}^* \neq 0$).  Coulson's
Fig.~5.7 is a graph of $\gamma(q)~vs.~q$, which clearly shows the
non-analytical behavior, although he did not comment on
it.\cite{Coulson} The non-analytical behavior of the energy as a
function of $q$ is also seen in GC-DFT.\cite{PPLB} Eq.~(\ref{III.6})
embodies the EVB representation of that behavior.  The non-analytical
behavior results in a derivative discontinuity in the energy as a
function of charge at integer values of the charge.  Perhaps because
orthogonality between resonance states is assumed, Nalewajski does not
comment on the behavior of the derivative at integer
charges.\cite{Nal-KS} Cios{\l}owski and Stefanov do see their version
of the NDOL limit as connected with PPLB.\cite{CioStef} Likewise,
PPLB,\cite{PPLB} Perdew,\cite{PerdewNATO} and Cios{\l}owski and
Stefanov\cite{CioStef} note that the derivative discontinuity
disappears once the systems in the GC representation begin to interact
significantly.  Eq.~(\ref{III.5}) embodies that behavior as well.
Examples of these charge dependences are illustrated below.

Starting from Eq.~(\ref{III.6}), we evaluate $H_{ci}$ at the
optimum NDOL $q$,
\begin{equation}
     q_{\rm opt}^{\rm NDOL} = \frac{1}{2} \, \Big(1 - \frac{1}
     {\sqrt{1+\epsilon^2}} \Big) \ , \label{IV.12}
\end{equation}
where $\epsilon = 2 \, H_{ci}/(H_{ii} - H_{cc})$.  The ground state
corresponds to the negative root of $\gamma(q)$ in
Eq.~(\ref{III.6.1}), since as $H_{ci} \rightarrow 0$, $q$ must also go
to zero, assuming that $H_{ii} - H_{cc} > 0$.  For convenience, we call
the ground state value of Eq.~(\ref{III.6}) $E_{\rm gs}^{\rm NDOL}
= E^{\rm NDOL}(q_{\rm opt}^{\rm NDOL})$.  The solution for
the resonance energy is equivalent to the well-known EVB
expression\cite{Warshel}
\begin{eqnarray}
     |H_{ci}| = \sqrt{(H_{cc} - E_{\rm gs}^{\rm NDOL}) (H_{ii} -
     E_{\rm gs}^{\rm NDOL})} \ .  \nonumber \\
     \label{IV.13}
\end{eqnarray}
Substituting Eq.~(\ref{IV.13}) into Eq.~(\ref{III.6}), we find
\begin{widetext}
\begin{equation}
	E^{\rm NDOL}(q) = (1-q) \, H_{cc} - 2 \,
	\sqrt{(1-q)(H_{cc} - E_{\rm gs}^{\rm NDOL}) \, q (H_{ii} -
	E_{\rm gs}^{\rm NDOL})} + q \, H_{ii} \ .  \label{IV.14}
\end{equation}
\end{widetext}
The dependence of $E^{\rm NDOL}(q)$ on $H_{cc}$ and $H_{ii}$
appears to be different from that implied by Eq.~(\ref{III.5b}).  In
fact, by setting $E(q_0^{\rm NDOL}) = E_{\rm gs}^{\rm
NDOL}$ in Eq.~(\ref{III.5b}), the two expressions become identical.

\begin{figure}
     \includegraphics[scale=0.5]{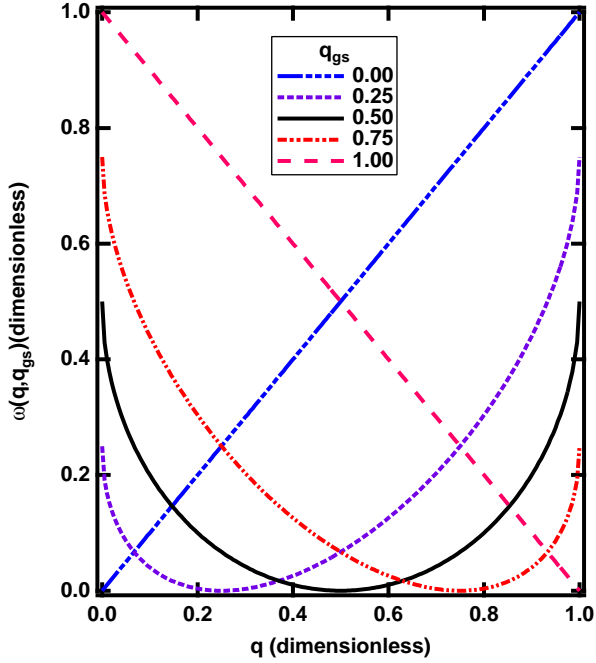}
     \caption{\label{fig:omega-NDOL} 2-state EVB-NDOL occupation number,
     Eq.~(\ref{IV.15e}), in the ensemble representation of $E^{\rm
     NDOL}(q)$, Eq.~(\ref{IV.15c}).  Charges are dimensionless.}
\end{figure}

We can gain further insight from Eq.~(\ref{IV.14}) by solving for
$H_{cc}$ and $H_{ii}$ in terms of $E_{\rm gs}^{\rm NDOL} \approx E_{\rm
gs}$, $E_{\rm xs}$, and $q_{\rm opt} \approx q_{\rm gs}$.  In a typical
diatomic, $E_{\rm xs}$ corresponds to the first electronic state which
dissociates to the ions ${\rm A^+}$ and ${\rm B^-}$.  Because of the
NDOL approximation, these three pieces of information are sufficient to
specify the energy.  To achieve the desired result, we first use the
fact that $H_{ci}^2$ can be derived from either eigenvalue to find that
$H_{ii} = E_{\rm xs} + E_{\rm gs} - H_{cc}$.  Next we solve for
$H_{cc}$ from Eq.~(\ref{IV.12}) assuming that we know $q_{\rm gs}$.
The result is that
\begin{equation}
     H_{cc} = (1 - q_{\rm gs}) \, E_{\rm gs} + q_{\rm gs} \, E_{\rm xs}
     \label{IV.15a}
\end{equation}
and
\begin{equation}
     H_{ii} = q_{\rm gs} \, E_{\rm gs} + (1 - q_{\rm gs}) \, E_{\rm xs}
     \ .
     \label{IV.15b}
\end{equation}
In the NDOL approximation, the pure state energies are simple linear
combinations of the eigenenergies.  Making all of the necessary
substitutions and rearrangements in Eq.~(\ref{IV.14}), we achieve the
ensemble representation:
\begin{eqnarray}
     E^{\rm NDOL}(q) &=& E_{\rm gs} + \omega(q,q_{\rm gs}) \,
     (E_{\rm xs} - E_{\rm gs})\ ,
     \label{IV.15c}
\end{eqnarray}
where the occupation number is
\begin{equation}
     \omega(q,q_{\rm gs}) = q_{\rm gs} - 2 \, \sqrt{q (1-q) q_{\rm gs}
     (1-q_{\rm gs})} + q \, (1 - 2 q_{\rm gs}) \ .
     \label{IV.15e}
\end{equation}
To see that Eq.~(\ref{IV.15c}) has the desired properties, note that
$\omega(q,q_{\rm gs})$ lies between 0 and 1 over the interval [0,1] in
$q$, it is 0 at $q = q_{\rm gs}$, and $\partial \omega(q,q_{\rm
gs})/\partial q$ at $q = q_{\rm gs}$ is also 0.  These properties are
illustrated in Fig.~\ref{fig:omega-NDOL}.  Taking into account our
restriction to a 2-state model, if we apply atom decomposition to the
eigenenergies, we would obtain the same form as Eq.~(\ref{I.1a}), with
$\omega_{\rm A^0} = 1 - \omega(q,q_{\rm gs})$, $\omega_{\rm A^+} =
\omega(q,q_{\rm gs})$, and $\omega_{\rm A^-} = 0$.  Importantly, the
representation is in terms of eigenenergies instead of energy
matrix elements.  Furthermore, as $q_{\rm gs}$ approaches 0, as in the
assumed dissociation limit for AB, $E^{\rm NDOL}(q)$ becomes linear in
$q$.

Another representation of the occupation number, Eq.~(\ref{IV.15e}), is
significant.  By using the relationship $q = \gamma^2/(1+\gamma^2)$
from Eqs.~(\ref{III.6.1}), we obtain
\begin{equation}
     \omega(\gamma,\gamma_{gs}) = \frac{(\gamma - \gamma_{\rm gs})^2}{(1
     + \gamma^2) \, (1 + \gamma_{\rm gs}^2)} \ .
     \label{IV.15f}
\end{equation}
As is necessary physically, the occupation number is 0 when $\gamma =
\gamma_{\rm gs}$, where $\gamma_{\rm gs}$ corresponds to $q_{gs}$.  The
complete generalization of Eq.~(\ref{IV.15f}) is equivalent to
following the process steps outlined above.

\begin{figure}
     \includegraphics[scale=0.5]{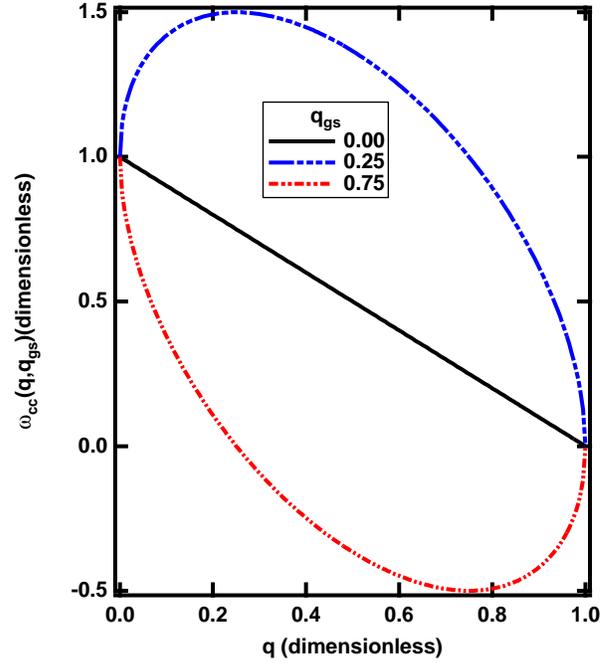}
     \caption{\label{fig:omega-H} 2-state EVB-NDOL coefficient for the
     covalent state given by Eq.~(\ref{IV.15h}).  The ionic state
     coefficient is obtained by reflecting this Figure through
     $\omega_{cc} = 1/2$.  Charges are dimensionless.}
\end{figure}

Eqs.~(\ref{IV.15a}) and (\ref{IV.15b}) can be inverted.  Inversion gives
$E^{\rm NDOL}(q)$ in terms of $H_{cc}$ and $H_{ii}$:
\begin{equation}
     E^{\rm NDOL}(q) = \omega_{cc} \, H_{cc} + \omega_{ii} \, H_{ii}
     \ , \label{IV.15g}
\end{equation}
where
\begin{equation}
     \omega_{cc} = (1 - \omega - q_{\rm gs})/(1 - 2 \, q_{\rm gs})
     \label{IV.15h}
\end{equation}
and
\begin{equation}
     \omega_{ii} = (\omega - q_{\rm gs})/(1 - 2 \, q_{\rm gs})
     \ . \label{IV.15i}
\end{equation}
Consistent with the conclusions of MM and physical necessity, the
coefficients are not positive-semidefinite.   Fig.~\ref{fig:omega-H}
illustrates Eq.~(\ref{IV.15h}), the coefficient for the covalent state.
In order to cover the range of energies between $E_{\rm gs}$ and
$E_{\rm xs}$, the coefficients of $H_{cc}$ and $H_{ii}$ cannot possibly
be positive-semidefinite.  The coefficients are not defined for $q_{gs}
= 1/2$.  At that value of the ground-state charge, $H_{cc}$ must equal
$H_{ii}$.

It is interesting to see how the present results connect with classical
electrostatic potentials (Eq.~(\ref{I.1})).  We can expand
Eq.~(\ref{IV.15e}) locally as a function of $q$ as long as the
expansion point is not zero or one.  First,
\begin{equation}
     \frac{\partial \omega(q,q_{\rm gs})}{\partial q} = -(1 -
     2 \, q) \, \sqrt{\frac{q_{\rm gs} \, (1-q_{\rm gs})}{q \, (1-q)}} +
     1 - 2 \, q_{\rm gs} \ .
     \label{IV.16}
\end{equation}
The obvious value about which to expand is $q_{\rm gs}$.  By
construction, $\partial \omega(q,q_{\rm gs})/\partial q
|_{q_{\rm gs}} = 0$.  Second,
\begin{equation}
     \frac{\partial^2 \omega(q,q_{\rm gs})}{\partial q^2} =
     \sqrt{\frac{q_{\rm gs} \, (1-q_{\rm gs})}{2 \, (q \, (1-q))^3}} \ .
     \label{IV.17}
\end{equation}
If we evaluate Eq.~(\ref{IV.17}) at $q = q_{\rm gs}$, as $q_{\rm gs}
\rightarrow$ 0 or 1, the expansion behaves badly.  Alternatively, we
might try expanding about $q = 1/2$, for the physically appealing
reason that $dE^{\rm NDOL}(q)/dq |_{q=1/2} = {\rm IP}_{\rm A}^* - {\rm
EA}_{\rm B}^*$.  Then
\begin{equation}
     \frac{\partial^2 \omega(q,q_{\rm gs})}{\partial q^2}
     \Bigg|_{q=1/2} = \sqrt{2 \, q_{\rm gs} \, (1-q_{\rm gs})} \ .
     \label{IV.18}
\end{equation}
To second order,
\begin{eqnarray}
     E^{\rm NDOL}(q) &\approx& E_{\rm gs} \nonumber \\
     &+& \Big(q_{\rm gs} - (1-\sqrt{2}/8) \sqrt{q_{\rm gs} \, (1-q_{\rm
     gs})} \nonumber \\
     &+& (1 - 2 \, q_{\rm gs} - \sqrt{q_{\rm gs} \, (1-q_{\rm gs})/2})
     \, q \nonumber \\
     &+& \sqrt{2 \, q_{\rm gs} \, (1-q_{\rm gs})} \ q^2/2 \Big) \,
     (E_{\rm xs} - E_{\rm gs}) \ .  \nonumber \\
     \label{IV.19}
\end{eqnarray}
This quadratic expansion has the form of Eq.~(\ref{I.1}) and behaves
well physically under dissociation, even to the extent of preserving
the dissociation limit.  The electronegativity, hardness, and
electrostatic contributions are embedded in the eigenenergies.  The
equivalent expansion of Eq.~(\ref{IV.15g}) might be more revealing in
displaying these contributions.  A key observation is that the
quadratic term vanishes completely at infinite separation ($q_{\rm gs}
\rightarrow$ 0) and only the linear dependence survives.  Again, the
survival of the linear charge dependence is consistent with PPLB. There
is no residual atomic hardness contribution as has appeared in many
implementations of Eq.~(\ref{I.1}).\cite{RSB,QEq,ES+,ToddM,abinitFF} We
speculate that a hardness contribution might be missing because we have
considered only a 2-state model instead of a 3-state model.  Our
reasoning behind this speculation is discussed below.  We have ruled
out the possibility that introducing the NDOL approximation prior to
making the expansion is a factor.  Nevertheless, adding a third state
must not change the fact that the coefficient of the quadratic term
must go zero in order for the results to be consistent with
PPLB.\cite{PPLB,CioStef,Nal-KS} Likewise, any hardness contributions to
the linear term must be scaled by a coefficient that goes to zero at
large $R$.  Clearly, one cannot approximate either Eq.~(\ref{III.5}) or
(\ref{III.6}) quadratically to arbritrary accuracy in a global sense.

\subsection{General Definition of the Energy \\ for a Given Charge}
Next we consider the densities and their energies in the NDOL
approximation.  From Eq.~(\ref{III.1}) we have
\begin{eqnarray}
      \rho^{\rm NDOL}(\gamma) = (\rho_{cc} + \gamma^2 \, \rho_{ii})/(1
      + \gamma^2) \ ,
      \label{IV.20}
\end{eqnarray}
Thus, from Eq.~(\ref{III.6.1}),
\begin{eqnarray}
     \rho^{\rm NDOL}(q) = \rho_{cc} + q \, (\rho_{ii} - \rho_{cc}) \ .
     \label{IV.22}
\end{eqnarray}
By symmetry in Eq.~(\ref{IV.20}), $\rho(\gamma) = \rho(-\gamma)$,
but, from Eq.~(\ref{II.5}), $E(\gamma) \neq E(-\gamma)$.  To
address this difficulty, MM appeal to a maximum entropy principle.  We
prefer instead to appeal to CS-DFT.\cite{Levy,SMV,Nal-KS} Accordingly,
CS-DFT instructs us to place the $\Gamma^{\rm NDOL}(\gamma)$ into
groups defined by the density that they produce.  Since $\Gamma^{\rm
NDOL}(\gamma)$ and $\Gamma^{\rm NDOL}(-\gamma)$ both yield the
same density, $\rho^{\rm NDOL}(q)$, they are grouped together.  The
energy assigned to $E^{\rm NDOL}[\rho^{\rm NDOL}(q)]$ is the minimum of
the energies for these two $\Gamma$'s:
\begin{eqnarray}
      E^{\rm NDOL}[\rho^{\rm NDOL}(q)] = \min \{ \langle H,\Gamma^{\rm
      NDOL}(\gamma(q))\rangle, \nonumber \\
      \langle H,\Gamma^{\rm NDOL}(-\gamma(q))\rangle\} \ .
      \label{IV.23}
\end{eqnarray}
The ground state energy is then the minimum over all $q$ of $E^{\rm
NDOL}[\rho^{\rm NDOL}(q)]$.  When differential overlap ($\rho_{ci}
\neq 0$) is included, the densities appear to become unique for
$\gamma$'s of different signs.\cite{enote1}

Note that a similar procedure could be followed for MM's 3-state case
at the ZDO level that they assume.  See also Ref.~\onlinecite{Nal-KS}.
Even with this comparatively simple extension, the situation is less
clear than in the 2-state case.  Many densities may have the same
charge on the atoms.  By again appealing to CS-DFT, one can assign
$E(q)$ for a given $q$ by minimizing over the energies of all densities
with the same $q$.  That is, as a straightforward extension of CS-DFT,
we very generally define
\begin{equation}
      E(q) \equiv \min_{\rho(q,\bm{\delta}) \rightarrow q} E^{\rm
      CS-DFT}[\rho(q,\bm{\delta})] \ ,
      \label{IV.24}
\end{equation}
where $\bm{\delta}$ represents all of the other undetermined parameters
of the density $\rho(q,\bm{\delta})$.  By minimizing over
$\bm{\delta}$, one may introduce dependencies on the energy matrix
elements that are absent from the present 2-state model.\cite{Nal-KS}
This is the reason behind our speculation that the isolated-atom
hardnesses do not appear in Eq.~(\ref{IV.19}) because of the
limitations of the 2-state model.  One advantage of the
charge-generalized CS-DFT approach over a maximum entropy principle is
that the true ground state properties can be preserved in the same way
that the ground state energy can be recovered {\textemdash} by
minimizing over all allowed densities in conventional
CS-DFT.\cite{Levy}

\subsection{Examples: HF, LiH, and $\bm {{\rm H_2}}$}
First we discuss modeling the polar molecules HF and LiH in the NDOL
approximation.  Then, for the nonpolar molecule $\rm {H_2}$, we compare
the NDOL and general cases.  There we utilize the Weinbaum
wavefunction.\cite{Weinbaum,CoulFisch} It provides an excellent
illustration of the ambiguities encoountered in defining the resonance
state wavefunctions.  We also examine various approximations for
$\delta N_{ci,\rm A}^{*}$ and examine the change in the charge
dependence as a function of $R$.

For HF and LiH, we use RKR curves\cite{RKR} to define $E_{\rm gs}$ and
$E_{\rm xs}$.  For the X$^1\Sigma^+$ and B$^1\Sigma^+$ states of HF,
the RKR data are from Di Lonardo and Douglas,\cite{DiLDoug}; for the
X$^1\Sigma^+$ and A$^1\Sigma^+$ states of LiH, the RKR data are from
Chan {\it et al.}\cite{Chan} and Pardo {\it et al.}\cite{Pardo} We use
the calculations of Ref.~\onlinecite{CioStef} to define $q_{\rm gs}$.
These data are shown in Figs.~\ref{fig:HF-fit} and \ref{fig:LiH-fit}.
The B$^1\Sigma^+$ and A$^1\Sigma^+$ states dissociate to ions, H$^+$
and F$^-$ and Li$^+$ and H$^-$, respectively.  To allow matching of the
different spatial ranges of the data, analytical fits for the energy
curves were made with the Rose and Rose+ionic functional
forms.\cite{Rose} The charge data were fit with the functional form,
$K_0+(K_1-K_0) (q^{K_2}/(1+(q^{K_2}+K_3^{K_2})))$, where the $K$'s are
fitting parameters.  $H_{cc}$ and $H_{ii}$ were computed from
Eqs.~(\ref{IV.15a}) and (\ref{IV.15b}), respectively.  In both cases,
$H_{cc}$ and $H_{ii}$ cross at $q_{\rm gs} = 1/2$ and meet their
respective states at the dissociation limits.  Representative shapes of
the charge dependence for given values of $R$ are shown in
Figs.~\ref{fig:HF-cuts} and \ref{fig:LiH-cuts}.  The correct
dissociation-limit behavior is observed in both cases.  In simulations
where each atom remains within a unit charge interval, the NDOL model
might therefore prove to be useful, although it is unlikely to be
quantitative.  The difference in charge transfer characteristics
between the present results and PPLB can be seen by contrasting the $R
= 3$ {\AA} curve of Fig.~\ref{fig:LiH-cuts} with the $R_c = 3.1$ {\AA}
curve of their Fig.~1.\cite{PPLB} The transfer is more gradual here and
passes through fractional charge states, compared to PPLB which is very
sharp and passes directly from a completely covalent state to a
completely ionic one.

Next we consider $\rm {H_2}$.  The simplest valence bond form for the
covalent state $\psi_c^{(1)} = (\phi_{\rm A}(1)\phi_{\rm
B}(2)+\phi_{\rm B}(1)\phi_{\rm A}(2))/(2+2 S_{\rm AB}^2)^{1/2}$, where
$S_{\rm AB}$ is the atomic orbital overlap.\cite{Slater} Our na\"{i}ve
inclination for the ionic state is to use the familiar $\psi_i^{(1)} =
(\phi_{\rm A}(1)\phi_{\rm A}(2) + \phi_{\rm B}(1)\phi_{\rm B}(2))/(2 +
2 S_{\rm AB}^2)^{1/2}$.  The total wavefunction is $\psi = c(\psi_c +
\gamma \psi_i)$.  It turns out that the total densities for both states
are identical: $\rho_{cc} = \rho_{ii} = (\phi_{\rm A}^2 + 2 \, S_{\rm
AB} \, \phi_{\rm A}\phi_{\rm B} + \phi_{\rm B}^2)/(1 + S_{\rm AB}^2)$.
(Designation of the electronic coordinate is suppressed in the
densities for readability.)  The interference density is $\rho_{ci} =
((\phi_{\rm A}^2+\phi_{\rm B}^2) \, S_{\rm AB}^2 + 2 \, \phi_{\rm
A}\phi_{\rm B})/(1+ S_{\rm AB}^2)$.

\begin{figure}
     \includegraphics[scale=0.333]{HF_fit_graph.EPSF}
     \caption{\label{fig:HF-fit} EVB model of HF. All energies are in
     eV, distances in {\AA}, and charges dimensionless.  RKR data from
     Ref.~\onlinecite{DiLDoug}.  Charges from
     Ref.~\onlinecite{CioStef}.}
\end{figure}

\begin{figure}
     \includegraphics[scale=0.333]{LiH_fit_graph.EPSF}
     \caption{\label{fig:LiH-fit} EVB model of HF. All energies are in
     eV, distances in {\AA}, and charges dimensionless.  RKR data from
     Ref.~\onlinecite{Chan} and \onlinecite{Pardo}.  Charges from
     Ref.~\onlinecite{CioStef}.}
\end{figure}

\begin{figure}
     \includegraphics[scale=0.45]{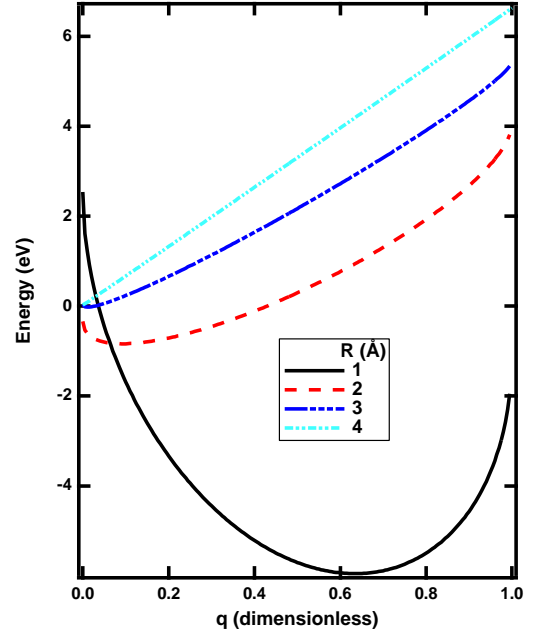}
     \caption{\label{fig:HF-cuts} Eq.~(\ref{IV.15c}) at discrete values
     of $R$ for HF. All energies are in eV, distances in {\AA}, and
     charges dimensionless.}
\end{figure}

\begin{figure}
     \includegraphics[scale=0.45]{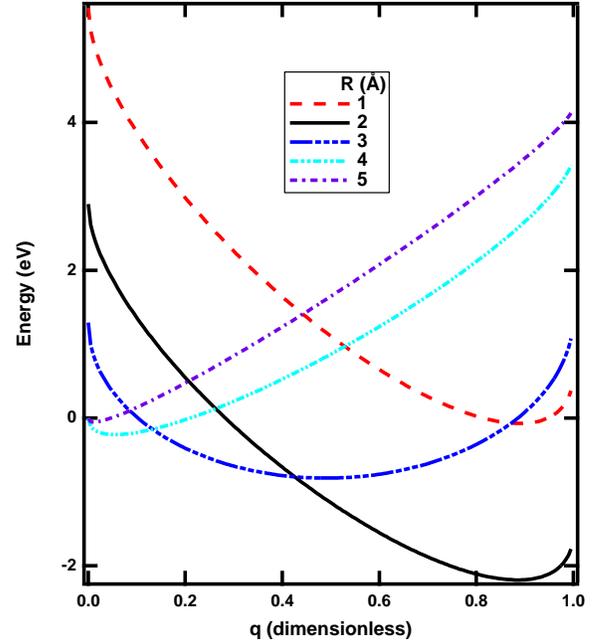}
     \caption{\label{fig:LiH-cuts} Eq.~(\ref{IV.15c}) at discrete values
     of $R$ for LiH. All energies are in eV, distances in {\AA}, and
     charges dimensionless.}
\end{figure}

If we assume a simple Hirshfeld partitioning as our total density
decomposition strategy,\cite{hir,NalParr2,NalParr3} then it is natural
to assume that each component of the density is likewise scaled by
$\phi_{\rm A}^2/(\phi_{\rm A}^2+\phi_{\rm B}^2)$ in order to obtain the
atom A contribution.  However, we find that because of symmetry,
$\delta N_{ci,\rm A}^{*} = 0$, which cannot be correct.  Of course, the
problem is that the assumption about the Hirshfeld partitioning form is
incorrect.  This partitioning cannot lead to a value of $\rho_{ii,{\rm
A}}$ that integrates to ${N_{\rm A^0}-1} = 0$.  A different
partitioning must be chosen to force all of the density in $\rho_{ii}$
to belong to atom B. In addition, we know the optimum values of
$\gamma$ for these two resonance states\cite{Weinbaum,CoulFisch} and
that these values correspond to $q = 0$.  For instance, at the
equilibrium $\rm {H_2}$ separation, $\gamma_{\rm opt} = -0.26$, with
the resonance energy defined as positive.\cite{Weinbaum} From
Eq.~(\ref{III.3}), we also can deduce that $\delta N_{ci,\rm A}^{*} =
\gamma_{\rm opt}/2 = -0.13$.  This analysis hints at the subtle
properties that the density decomposition must possess if one
implements explicit decomposition of $\rho_{ci}$.

Another immediate insight is that any fixed value of $\gamma$ applied
within $\psi$ leads to an acceptable covalent state.  Clearly, the
energy change caused by a polarization of the $\rm {H_2}$ charge
density will differ depending on one's choice for the resonance state.
Similarly, the density decomposition will show some sensitivity to this
choice.

Of course, a more sensible choice for the ionic state is $\psi_i^{(2)} =
\phi_{\rm B}(1)\phi_{\rm B}(2)$ because, for a homonuclear diatomic,
the charged states correspond to broken charge-symmetry states.  Even
in this simple case, the partitioning of $\rho_{ci}$ may be nontrivial.
% For simplicity, we retain $\psi_c = (\phi_{\rm A}(1)\phi_{\rm
% B}(2)+\phi_{\rm B}(1)\phi_{\rm A}(2))/(2+2 S_{\rm AB}^2)^{1/2}$ as the
% covalent wavefunction.

\begin{figure}
     \includegraphics[scale=0.50]{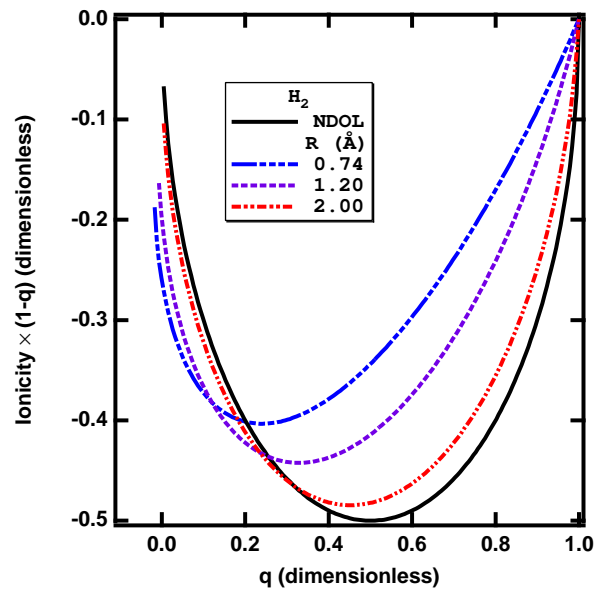}
     \caption{\label{fig:H2-ion} Ionicity scaled by $1-q$ at discrete
     values of $R$ for $\rm {H_2}$.  Distances are in {\AA} and charges
     are dimensionless.}
\end{figure}

As a final consideration, we illustrate the influence of overlap on
ionicity.  For this purpose, we utilize our original choices for
covalent and ionic wavefunctions, $\psi_c^{(1)}$ and $\psi_i^{(1)}$.
The overlap integral is estimated in terms of $S_{\rm AB}$ as given
above.  Taking $\phi(\bf r) = \sqrt{\alpha^3 /\pi} \, {\rm exp}(-\alpha
\, r)$, the atomic overlap is $S_{\rm AB}(a) = {\rm exp}(-a) (1 + a +
1/3 \, a^2)$, where $a = \alpha \, R$.  The effect of overlap is shown
in Fig.~\ref{fig:H2-ion} as $\gamma(q) (1-q)$.  There it can been seen
that the NDOL approximation becomes accurate beyond approximately 2
{\AA}.  The relatively large range of $R$ for which the NDOL
approximation is accurate in this case is due in no small part to the
fact that $S_{ci}$ depends on the square of $S_{\rm AB}$.  This broad
range of accuracy and the fact that Eqs.~(\ref{IV.15c}) and
(\ref{IV.15e}) are well-behaved for any $R$ makes it tempting to use
NDOL in general.  However, near equilibrium bond lengths,
Fig.~\ref{fig:H2-ion} indicates that overlap effects should not be
ignored.

\section{Conclusion}
In order to apply chemical potential equalization in a simulation that
involves conditions far from reference states, the potential energy
must be defined for arbitrary values of the charges.  To address this
challenge, we have derived a new charge-dependent pair potential from a
2-state empirical valence bond model.  The charge is defined from a
decomposition of the density into constituent contributions.  The
explicit charge dependence is deduced by requiring consistency between
the density decomposition and the wavefunction descriptions of the
ground state.  The energy expression can be made valid for any range of
charge of interest.  The decomposition theme is further extended to
define the energy of individual constituents, again by requiring
consistency between the density decomposition and the wavefunction
descriptions.  The energy of the system for a given value of charge is
made unique by appealing to constrained search density functional
theory.  An examination of the model shows linear dependence on charge
at the dissociation limit, as well as discontinuous behavior in the
derivative of the energy as a function of charge at integer values of
the charge.  This behavior is consistent with the analysis of Perdew
{\it et al.}\cite {PPLB} and others.\cite{CioStef,Nal-KS} The potential
energy for arbitrary charge and separation of the constituents can be
represented as an ensemble average of the eigenenergies with a
nonlinear, analytical dependence of the occupation number on charge.
The representation of the potential energy in terms of the pure-state
energy matrix elements is possible, but, by physical necessity, cannot
be expressed as an ensemble average with positive semidefinite
coefficients.

To determine a pair potential for all values of $q$ and $R$ with this
method, one needs five reference or calibration curves.  These most
often will be computed values of the ground-state energy, the charges
along the ground-state energy curve, the resonance state overlap
integral, the covalent energy and ionic energy.  Measurements of ground
and ionic-excited state energies and charges can also be used.  The
covalent energy and ionic energy are determined for integer charges
only.  To determine a point on the potential energy curve, a
well-defined five step procedure is followed using the five input
curves.  An immediate application of the method could be to construct
the reference potential curves for the $\rm A_2$, $\rm B_2$, and AB
systems.  Simulations on arbitrary mixtures of these three types of
systems under nonequilibrium initial conditions would then be possible.
Chemical potential equalization would be used to dynamically adjust the
charges of the constituents.  More generally, we envision the
present approach as forming the basis for a new class of
charge-dependent empirical potentials for use in large-scale
simulations of reactive systems.

\begin{acknowledgments}
The work of S.M.V.~was performed in part at Los Alamos National
Laboratory under the auspices of the U.~S.~Department of Energy, under
contract No.~W-7405-ENG-36, and funded through its Center for
Semiconductor Modeling and Simulation, a CRADA program performed
jointly with the Semiconductor Research Corporation, and through the
Advanced Fuel Cycle Initiative.  S.M.V.~thanks the University of New
Mexico, Department of Physics and Astronomy for its hospitality during
the 2003-2004 academic year.  S.R.A.~would like to thank the National
Science Foundation for support during the initial stages of this work
(DMR-9520371).  This work was supported by National Science
Foundation grant No.~CHE-0304710.
\end{acknowledgments}


\begin{thebibliography}{99}
\bibitem{enote1}{For an arbitrary set of basis wavefunctions, there is
no simple way to determine which wavefunctions map to which densities.
Even in a finite basis set, pathological cases are possible.}
\end{thebibliography}
\end{document}